\input harvmac \noblackbox
\def\ev#1{\langle#1\rangle}

\def\A{{\cal A}}
\def\fund{  \> {\vcenter  {\vbox
              {\hrule height.6pt
               \hbox {\vrule width.6pt  height5pt
                      \kern5pt
                      \vrule width.6pt  height5pt }
               \hrule height.6pt}
                         }
                   } \>
           }
%

\nref\DouglasBA{ M.~R.~Douglas and N.~A.~Nekrasov,
hep-th/0106048.
}

\nref\Ardalan{F. Ardalan and N. Sadooghi, hep-th/0002143 Int. J.
Mod. Phys. A. (2001); hep-th/0009233.}

\nref\Gracia-BondiaPZ{ J.~M.~Gracia-Bondia and C.~P.~Martin,
Phys.\ Lett.\ B {\bf 479}, 321 (2000) [hep-th/0002171].
}

\nref\MartinQF{ C.~P.~Martin,
hep-th/0008126.
}

\nref\BonoraHE{ L.~Bonora, M.~Schnabl and A.~Tomasiello,
Phys.\ Lett.\ B {\bf 485}, 311 (2000) [hep-th/0002210].
}

\def\anomrefs{\refs{\Ardalan - \BonoraHE}}
\def\Martin{\MartinQF}

\lref\SeibergVS{ N.~Seiberg and E.~Witten,
JHEP {\bf 9909}, 032 (1999) [hep-th/9908142].
}
\def\SW{\SeibergVS}
\lref\CallanSA{ C.~G.~Callan and J.~A.~Harvey,
Nucl.\ Phys.\ B {\bf 250}, 427 (1985).
} \lref\BlumYD{ J.~D.~Blum and J.~A.~Harvey,
Nucl.\ Phys.\ B {\bf 416}, 119 (1994) [hep-th/9310035].
}

\lref\GrossSS{ D.~J.~Gross and N.~A.~Nekrasov,
JHEP {\bf 0103}, 044 (2001) [hep-th/0010090].
}
\lref\NekrasovSS{ N.~Nekrasov and A.~Schwarz,
dimensional theory,'' Commun.\ Math.\ Phys.\  {\bf 198}, 689
(1998) [hep-th/9802068].
}
\lref\DouglasSW{ M.~R.~Douglas and G.~Moore,
hep-th/9603167.
} \lref\JBKI{J.~D.~Blum and K.~Intriligator,
Nucl.\ Phys.\ B {\bf 506}, 223 (1997) [hep-th/9705030];
Nucl.\ Phys.\ B {\bf 506}, 199 (1997) [hep-th/9705044].
}

\lref\LiuPK{ H.~Liu and J.~Michelson,
hep-th/0104139; hep-th/0107172.
}
\lref\LiuMJ{ H.~Liu,
hep-th/0011125.
}
\lref\LiuAD{ H.~Liu and J.~Michelson,
hep-th/0008205.
}

\lref\LiuQH{ H.~Liu and J.~Michelson,
Phys.\ Rev.\ D {\bf 62}, 066003 (2000) [hep-th/0004013].
}

\lref\GomisBN{ J.~Gomis, M.~Kleban, T.~Mehen, M.~Rangamani and
S.~Shenker,
JHEP {\bf 0008}, 011 (2000) [hep-th/0003215].
}
\lref\OkawaMV{ Y.~Okawa and H.~Ooguri,
hep-th/0104036.
}

\lref\MinwallaPX{ S.~Minwalla, M.~Van Raamsdonk and N.~Seiberg,
JHEP {\bf 0002}, 020 (2000) [hep-th/9912072].
}

\lref\VanRaamsdonkRR{ M.~Van Raamsdonk and N.~Seiberg,
JHEP {\bf 0003}, 035 (2000) [hep-th/0002186].
}

\lref\IbanezQP{ L.~E.~Ibanez, R.~Rabadan and A.~M.~Uranga,
Nucl.\ Phys.\ B {\bf 542}, 112 (1999) [hep-th/9808139].
}

\lref\RajaramanDW{ A.~Rajaraman and M.~Rozali,
JHEP {\bf 0004}, 033 (2000) [hep-th/0003227].
}

\lref\LeighHJ{ R.~G.~Leigh and M.~Rozali,
Phys.\ Rev.\ D {\bf 59}, 026004 (1999) [hep-th/9807082].
}

\lref\LawrenceJA{ A.~E.~Lawrence, N.~Nekrasov and C.~Vafa,
Nucl.\ Phys.\ B {\bf 533}, 199 (1998) [hep-th/9803015].
}

\lref\MukhiVX{ S.~Mukhi and N.~V.~Suryanarayana,
JHEP {\bf 0105}, 023 (2001) [hep-th/0104045].
}

\lref\GarousiCH{ M.~R.~Garousi,
Nucl.\ Phys.\ B {\bf 579}, 209 (2000) [hep-th/9909214].
}

\Title{\vbox{\baselineskip12pt\hbox{hep-th/0107199}
\hbox{UCSD-PTH-01-14} }}
{\vbox{\centerline{$\star$-Wars Episode I: The Phantom Anomaly}}}
\centerline{ Ken Intriligator and Jason Kumar}
\bigskip
\centerline{Department of Physics} \centerline{University of
California, San Diego} \centerline{La Jolla, CA 92093-0354, USA}

\bigskip
\noindent

As pointed out, chiral non-commutative theories exist, and
examples can be constructed via string theory.  Gauge anomalies
require the matter content of individual gauge group factors,
including $U(1)$ factors, to be non-chiral.  All ``bad'' mixed
gauge anomalies, and also all ``good'' (e.g. for $\pi
^0\rightarrow \gamma \gamma$) ABJ type flavor anomalies, {\it
automatically} vanish in non-commutative gauge theories.  We
interpret this as being analogous to string theory, and an example
of UV/IR mixing: non-commutative gauge theories automatically
contain ``closed string,'' Green-Schwarz fields, which cancel
these anomalies.

\Date{July 2001}

\newsec{Introduction}

Quantum field theories on non-commutative spaces are a subject of
much recent interest, initiated by the realization that such
theories actually occur in the worldvolume of D-branes in a
background $B_{NS}$ field.  There are many fascinating
interconnections between non-commutative field theories and string
theory. See \DouglasBA\ for a recent review and references.  There
have even been speculations that perhaps nature is actually
noncommutative on some short distance scale.

We will here consider aspects of chiral \foot{By ``chiral
theories,'' we mean that there are chiral fermions in a
representation of the gauge group which is not vector-like.  Thus
e.g. a 4d $U(N)$ theory with a chiral fermion in the adjoint
representation will not be considered to be a chiral theory.  Such
fermions, in vector-like representations, can always be given a
large mass $\Tr~ m\psi _\alpha \psi ^\alpha$ and decoupled from
the theory, and thus never contribute to anomalies.  One might
worry about such decoupling arguments in the non-commutative
context, but it was explicitly verified in noncommutative $U(N)$
(see e.g. \Martin) that vector-like fermion representations indeed
do not contribute to anomalies.} non-commutative gauge theories
and anomaly cancellation. Non-commutative gauge theory anomalies
have been analyzed in several works \anomrefs, with the result
that the only possible anomaly-free matter content for a single
non-commutative gauge group is non-chiral. This is perhaps not
surprising, given that the gauge group is $U(N)$, with only
fundamental, anti-fundamental, and adjoint representations allowed
(e.g. $SU(5)$ with a ${\bf 5}+\bf{\overline{10}}$ is not allowed).

There are, however, consistent chiral non-commutative gauge
theories, based on product groups $\prod _i U(N_i)$.  Vanishing
$\Tr U(N_i)^3$ gauge anomalies constrain the matter content of any
given $U(N_i)$ gauge group to be non-chiral.  But, as with the
standard model, the representations of $\prod _i U(N_i)$ can be
chiral via some bi-fundamental chiral fermion matter
representations $(\fund _i , \overline{\fund _j})$ for which there
is no conjugate $(\overline {\fund _i}, \fund _j)$ chiral fermion.
Such chiral non-commutative gauge theories can be explicitly
constructed via D-branes at orbifold singularities.  As long as
the orbifold acts trivially on $B_{NS}$, there is no obstruction
to including a background $B_{NS}$ and making the world-volume
gauge theory non-commutative. Thus one can consider general
orbifolds, but not orientifolds.

A concrete example of a chiral non-commutative gauge theory which
exists via a string theory construction is that of $N$ D3 branes
at a $C^3/Z_3$ orbifold singularity.  The world-volume gauge
theory has ${\cal N}=1$ supersymmetry with the quiver gauge group
and chiral superfield matter content \eqn\stdmdl{\matrix{ \quad
&U(N)_1\times U(N)_2\times U(N)_3\cr X_i&(\fund ,
\overline{\fund}, 1)\cr Y_i&(1, \fund , \overline{\fund} )\cr
Z_i&(\overline{\fund} , 1, \fund ).\cr}} with $i=1,2,3$. As will
always be the case in what follows, the overall $U(1)$ is such
that the fundamental $\fund$ has charge $+1$, and the
anti-fundamental $\overline{\fund}$ has charge $-1$. There is no
obstruction to turning on a background $B_{NS}$, so the above
theory also must exist as a chiral non-commutative gauge theory.
One can similarly construct more general orbifold examples, e.g.
based on IIB D3 branes at a $C^3/(Z_N\times Z_M)$ singularity, or
even non-supersymmetric examples via general orbifolds.  As long
as $B_{NS}$ is not projected out, one can turn on its expectation
value and make the worldvolume theory noncommutative.

Commutative gauge theories can have both $\Tr U(N_i)^3$ anomalies
and also mixed anomalies, like $\Tr U(N_i)U(N_i)^2$; these mixed
anomalies can generally be non-zero, and they must be cancelled by
a generalized Green-Schwarz mechanism in order to obtain a
consistent theory.  (E.g. this is the case in the commutative
version of the theory \stdmdl.)  Note that the $\Tr U(N_i)^3$
anomaly comes entirely from planar diagrams, whereas the mixed
$\Tr U(N_i)U(N_j)^2$ anomaly comes entirely from non-planar
diagrams.  This is simply because the bi-fundamentals running in
the loop $(\fund _i , \overline{\fund _j})$ can be regarded as an
open string with $U(N_i)$ Chan-Paton factor at one end and
$U(N_j)$ at the other, so all $U(N_i)$ currents couple to one edge
of the annulus and all $U(N_j)$ currents to the other.

Because the mixed anomalies come from non-planar diagrams, they're
significantly altered by the non-commutativity.  A standard
explanation for anomalies is that one cannot regulate UV divergent
loop momentum integrals in a way which is compatible with the
naive, non-anomalous Ward identity.  For example, one would
recover the naive Ward identity if not for the fact that one
cannot freely shift the integrated loop momentum, as the integral
is linearly divergent in the UV. But, in the non-commutative case,
the loop-momentum dependent phase factors of non-planar diagrams
{\it do} regulate the loop integral in a way which is compatible
with the naive Ward identity. This can be seen explicitly in the
anomaly analysis of \Martin\ for a bifundamental chiral fermion:
the momentum loop integral is finite, and thus the loop momentum
shift, which leads to the naive Ward identities, is valid.  Thus
mixed anomalies {\it automatically} vanish in non-commutative
gauge theories.

The automatic vanishing of the mixed $\Tr ~U(N_i)\Tr ~U(N_j)^2$
occurs even if the $U(N_i)$ is a global symmetry (which can be
regarded as the $g_i\rightarrow 0$ limit of a gauge symmetry).
Thus both the ``bad'' mixed gauge anomalies, which would have
rendered the theory inconsistent, and the ``good'' (e.g. for  $\pi
^0\rightarrow \gamma \gamma$) ABJ type flavor anomalies
automatically vanish in the non-commutative theory.

\lref\GreenMN{ M.~B.~Green, J.~H.~Schwarz and E.~Witten, {\it
Superstring Theory. Vol. 2: Loop Amplitudes, Anomalies And
Phenomenology,} Cambridge, Uk: Univ. Pr. (1987).}
\lref\GreenSG{ M.~B.~Green and J.~H.~Schwarz,
Phys.\ Lett.\ B {\bf 149}, 117 (1984).
}

\lref\SheikhJabbariVM{ M.~M.~Sheikh-Jabbari,
Phys.\ Lett.\ B {\bf 455}, 129 (1999) [hep-th/9901080].
}

\lref\BigattiIZ{ D.~Bigatti and L.~Susskind,
Phys.\ Rev.\ D {\bf 62}, 066004 (2000) [hep-th/9908056].
}
\lref\YinBA{ Z.~Yin,
Phys.\ Lett.\ B {\bf 466}, 234 (1999) [hep-th/9908152].
}

This automatic vanishing of non-planar anomalies is reminiscent of
the situation in string theory, which originally led to the
Green-Schwarz mechanism \refs{\GreenSG, \GreenMN}.  From the
worldsheet viewpoint, an extra factor of $q^{p_{NP}^2/4}$, with
$p_{NP}$ the total non-planar momentum running through the
diagram, regulates the non-planar hexagon diagram; thus the mixed
anomaly vanishes.  On the other hand, a low-energy space-time
analysis of the field theory would seem to give a non-zero mixed
anomaly; the resolution of \GreenSG\ is that string theory
automatically cancels this mixed anomaly via a tree-level
contribution from a closed string state ($B_{\mu \nu}$). In the
closed string channel picture of the non-planar diagram, the
$q^{p_{NP}^2/4}$ factor can be considered as coming from poles of
these closed string states.

We interpret the automatically vanishing mixed anomalies of
non-commutative gauge theories as a UV/IR mixing effect, similar
to string theory: these theories automatically contain ``closed
string'' fields, with axionic coupling to the gauge fields, which
implement a Green-Schwarz anomaly cancellation mechanism.  The
cancelling of the mixed flavor anomalies in non-commutative gauge
theory is also analogous to string theory, which is believed to
have no global symmetries; any symmetry must be a gauge symmetry,
and therefore must be non-anomalous.

In the context of string theory constructions of chiral gauge
theories via branes at orbifold singularities, the automatic
anomaly cancellation is perhaps not so surprising.  It is already
known \refs{\DouglasSW, \JBKI, \LeighHJ, \IbanezQP} in the
commutative case that certain closed string modes (twisted RR
fields) remain present, even in the decoupling limit, and cancel
all mixed anomalies via a generalized Green-Schwarz mechanism;
e.g. this happens for the $C^3/Z_3$ example \stdmdl.  The
situation in the noncommutative case, to be discussed, is fairly
similar.

What is perhaps surprising is that one can apparently consider
quantum field theories not constructed via string theory, and thus
potentially sick due to uncancelled mixed anomalies.  Making space
noncommutative automatically cures these anomalies\foot{Some
string theorists believe that noncommutative quantum field
theories only exist as a limit of string theory.  The automatic
mixed anomaly cancellation perhaps bolsters this philosophy, with
the cancellation regarded as being simply inherited from the
underlying string theory.}.  Our interpretation is that the
noncommutative field theory, via UV/IR type effects, knows to
include the necessary ``closed string'' Green-Schwarz modes also
found in the string theory constructions.  We claim that these
``closed string'' fields $\chi _i$ couple to the noncommutative
$\prod _{i=1}^n U(N_i)$ gauge theory as \eqn\ncgsint{\sum _{i=1}^n
\int d^4 k \chi _i (k){\cal O}_{\chi _i}(-k),} with ${\cal
O}_{\chi _i} (-k)$ the gauge invariant analog of (the Fourier
transform of) $\Tr F_i \wedge F_i$, which involves an open Wilson
loop and is thus non-local.  If the $\chi _i(x)$ had constant
expectation value, they would give the theta angle terms of the
gauge groups, $\ev{\chi _i(x)}={\theta _i \over 2\pi}$ (see e.g.
\LawrenceJA).  However, the $\chi _i$ cannot be taken to be
constant. The mixed anomalies are cancelled because $\chi _i (k)$
are not exactly gauge invariant.  (The $\chi _i$ kinetic terms can
be taken to vanish, so they become non-dynamical in the decoupling
limit.)

Noncommutative gauge theories constructed via orbifolds were also
considered in \RajaramanDW, in the context of analyzing UV/IR
mixing of noncommutative field theories  \refs{\MinwallaPX,
\VanRaamsdonkRR}. In line with the proposal of \refs{\MinwallaPX,
\VanRaamsdonkRR}, it was argued in \RajaramanDW\ that UV
divergences associated with non-zero 1-loop beta functions,
involve closed string modes which couple to the noncommutative
gauge field kinetic terms.  Other works, e.g. \refs{ \LiuQH,
\GomisBN} have argued that the $p\circ p$ regulator
\refs{\MinwallaPX, \VanRaamsdonkRR}\ of non-planar diagrams are
better thought of as an open string effect, having to do with
stretched strings \refs{\SheikhJabbariVM, \BigattiIZ, \YinBA}. Of
course, in the string theory, there's some freedom as to whether
effects are due to open or closed strings, thanks to the
open-closed string channel duality.

In any case, the non-decoupled closed string modes discussed in
\RajaramanDW\ also come from the twisted closed string sector.  We
expect that these closed string modes $\phi _i$ come from twisted
sector NS fields, and that they couple as \eqn\ncgcing{\sum _i
\int d^4 k \phi _i (k){\cal O}_{\phi _i}(-k),} with ${\cal
O}_{\phi _i}(-k)$ the gauge invariant analog of $\Tr F_{\mu
\nu}F^{\mu \nu}$, which involves an open Wilson loop, and is thus
not local.  If the $\phi _i(x)$ had constant expectation value,
they would give the effective gauge coupling of the gauge groups,
$\ev{\phi _i(x)} = {4\pi i \over g_i^2}$ (see e.g. \LawrenceJA).
As discussed in \RajaramanDW, when the gauge groups have
non-vanishing 1-loop beta functions, the UV divergence can lead to
IR poles and a Wilsonian interpretation requires including the
fields $\phi _i$ in the theory. This is very similar to our
discussion above and claim that the closed string fields $\chi _i$
of \ncgsint\ are to be included in the theory.

In the commutative context, one can cancel all $U(1)$ anomalies
via a generalized Green-Schwarz mechanism.  This includes both
mixed anomalies and also anomalies in a single $U(1)$ factor with
chiral matter.  The noncommutative Green-Schwarz mechanism of
\ncgsint\ is capable of cancelling all mixed anomalies, but not
the $\Tr U(1)^3$ anomalies of a single, chiral $U(1)$ gauge
theory. We claim that, unlike the commutative version, $\Tr
U(1)^3$ anomalies of chiral noncommutative $U(1)$ theories can
{\it not} be cured by the Green-Schwarz mechanism.  Thus chiral
$U(1)$ theories do not admit noncommutative generalizations.  The
lack of a Green-Schwarz anomaly cancellation mechanism for chiral
non-commutative $U(1)$ could have been anticipated by its Morita
equivalence to non-commutative $U(N>1)$, as there is no
Green-Schwarz cancellation mechanism for commutative $U(N>1)^3$
anomalies.

One can also consider chiral gauge theories in other dimensions.
For example, consider $N$ IIB D5 branes at a $C^2/Z_2$
singularity.  The worldvolume gauge theory is a chiral 6d ${\cal
N}=(1,0)$ theory with quiver gauge group and hypermultiplet matter
content \DouglasSW \eqn\sixdthy{\matrix{ \quad &U(N)_1\times
U(N)_2\cr Q_1&(\fund , \overline{\fund})\cr Q_2&(\overline{\fund}
, \fund ).\cr}} The theory has a 6d ${\cal N}=(1,0)$ tensor
multiplet, which cancels the mixed anomalies \JBKI, and a
hypermultiplet which cancels \DouglasSW\ the anomalies associated
with the $U(1)$ with charged matter. The $\Tr SU(N_c)^4$
irreducible anomaly is proportional to $2N_c-N_f$ and thus
consistent theories must always have $N_f=2N_c$ for each $SU(N_c)$
factor, as is the case in \sixdthy. In the commutative context,
$U(1)$ factors with charged matter are always anomalous, and thus
generally get a mass in the process of anomaly cancellation
described in \DouglasSW.  In the non-commutative context, however,
it is the full irreducible $\Tr U(N_c)^4$ (rather than $\Tr
SU(N_c)^4$) which is proportional to $2N_c-N_f$, and this applies
for all $N_c$, including $N_c=1$.  So the $N_f=2N_c$ condition
(needed to cancel twisted sector tadpoles \refs{\DouglasSW,\JBKI})
found already in the commutative context ensures that the full
$\Tr U(N_c)^4$ anomalies cancel; no additional GS anomaly
cancellation is needed for the $U(1)$ factors, and the $U(1)$
factors do not get massive or decouple in the non-commutative
theory.

In section 2, we review Green-Schwarz anomaly cancellation in
commutative gauge theories.  In section 3, we discuss
noncommutative anomalies. In section 4 we discuss noncommutative
anomaly cancellation and noncommutative analogs of the
Green-Schwarz mechanism. In section 5 we comment on anomalies and
Wess-Zumino terms in Higgsed noncommutative gauge theories, with
an example similar to the standard-model.

\newsec{Commutative gauge anomalies, and Green-Schwarz cancellation}
The gauge anomaly of $d$ dimensional gauge theories is given by
\eqn\anomg{\delta _\lambda \log Z =2\pi i \int [\Tr _\rho \exp
({iF\over 2\pi}) ]^{(1)},} with $\rho$ the representations of the
chiral fermions and $I_{d+2}=dI^{(0)}$ and $\delta I^{(0)}=dI
^{(1)}$. In particular, in four spacetime dimensions the anomaly
comes from \eqn\anomp{\sim [\Tr _\rho (F)^3]^{(1)}\sim \Tr _\rho
(\lambda d(AdA+\half A^3)).} This anomaly is proportional to $\Tr
_\rho (\{T^a,T^b\}T^c)\equiv \A ~d^{abc}$, and the theory is sick
if this quantity is non-zero. For non-Abelian gauge groups, this
sickness cannot be cured.  (A cure is anomaly inflow of gauge
charge \refs{\CallanSA, \BlumYD} into extra dimensions, but our
interest is in $4d$ theories which decouple from any higher
dimensional bulk.  So gauge theories whose anomalies cannot be
cancelled purely in $4d$ will be considered as incurably sick.)

But abelian gauge theories with anomaly $\A ~\neq 0$ can always be
cured by a Green-Schwarz mechanism.  As usual, the condition for
Green-Schwarz anomaly cancellation is that the anomaly polynomial
$\Tr _\rho F^3$ should factorize.  The anomalies involving $U(1)$
groups trivially factorize, because we can drop the $\Tr$ for the
$U(1)$ factors.

For example, consider a theory with gauge group $U(1)$ and chiral
fermions with charges $q_i$.  The anomaly \anomp\ can be written
as \eqn\anomvi{\delta _\lambda \log Z \sim \A ~\int \lambda
d(AdA+\half A^3)= \A ~\int \lambda F\wedge F, \qquad \A ~\equiv
\sum q_i^3.} Note that generally \eqn\FFdual{\Tr F\wedge F=d\Tr
(AdA+{2\over 3}A^3),} but the $A^3$ term vanishes for commutative
$U(1)$ gauge theory, so the difference between the $\half$
coefficient in \anomp\ versus the ${2\over 3}$ in \FFdual\ is
immaterial.  The $A^3$ term does not vanish in the non-commutative
version of this theory, to be discussed in the following sections.

If $\A \neq 0$, the theory is sick, but can be cured by
introducing a scalar field ``axion'' $\chi$ with term in the
action \eqn\pff{S_{GS}=\alpha \int \chi ~ F\wedge F,} which
properly appears to be gauge invariant.  However, if it happens
that the gauge invariant field strength of $\chi $ is actually
\eqn\Hisi{H=d\chi + \beta  ~A,} then $\chi $ cannot be gauge
invariant, but must instead shift under gauge transformations,
\eqn\pgt{\delta _\lambda \chi =- \beta \lambda \quad \hbox{under}
\quad A\rightarrow A+d\lambda.}  The coupling \pff\ then leads to
tree-level anomaly contributions which cancels \anomvi\ provided
the constants $\alpha$ and $\beta$ satisfy \eqn\abcanc{\alpha
\beta =\A.} We can normalize $\chi$ so that $\alpha =1$ and $\beta
= \A$.  The gauge invariant $\chi $ kinetic term is (with $Z_\chi
$ a normalization constant) \eqn\gammass{-\half Z_\chi (\partial
_\mu \chi  +\beta ~A_\mu )^2,} which gives the photon a mass
$m_\gamma ^2=\beta ^2 Z_\chi $.  The cross term in \gammass\ can
be written as \eqn\crossi{-\beta Z_\chi B_2\wedge F,} where $B_2$
is the 2-form potential dual to $\chi $, $d\chi =*dB_2$ and we
integrated by parts.

Consider next a $U(1)\times G$ gauge theory. The anomaly can
contain terms \eqn\prodai{\delta_\lambda \log Z \sim \A _{111}
[F_1^3]^{(1)}+ \A _{1GG}[F_1F_GF_G]^{(1)}+\A
_{11G}[F_1F_1F_G]^{(1)}+\A _{GGG}[F_G^3]^{(1)},} with $\A
_{111}=\Tr _\rho U(1)^3$ and $\A _{11G}=\Tr _{\rho} U(1)^2G$ etc.
Suppose that $G$ is non-abelian, with no $U(1)$ factor, so $\A
_{11G}=0$, and that the $G$ matter content is chosen to be anomaly
free, so $\A _{GGG}=0$.  We can write the remaining anomaly in
\prodai\ as \eqn\prodaii{\delta _\lambda \log Z \sim \int \lambda
_1(\A _{111}F_1\wedge F_1+ \A _{1GG} \Tr F_G\wedge F_G).} This
anomaly can be cancelled via a field $\chi$ with coupling
\eqn\pffii{S_{GS}=\int \chi (\alpha _1 F_1\wedge F_1+\alpha _G \Tr
F_G\wedge F_G),} provided $\chi$ has gauge invariant field
strength $H=d\chi +\beta A_1$, so $\delta _\lambda \chi = -\beta
\lambda _1$, and the coupling constants satisfy $\alpha _1 \beta =
\A _{111}$ and $\alpha _G\beta =\A _{1GG}$.

The mixed anomaly $\A _{1GG}$ corresponds to a net $U(1)$ charge
of the fermion zero modes contained in the $G$ instanton's 't
Hooft vertex. If $\A _{1GG}$ is non-zero but cancelled by the
Green-Schwarz mechanism discussed above, the $G$ instanton t'
Hooft vertex is actually $U(1)$ neutral, as the Green-Schwarz
mechanism ensures that the $U(1)$ current is properly conserved.
The fermion zero modes still carry net $U(1)$ charge, but it is
screened by the factor of $\exp(i\alpha _G \chi )$ which
accompanies the instanton 't Hooft vertex because of \pffii.

More generally, to cancel all anomalies involving $U(1)_i$ factors
of a general gauge theory, one needs the same number of $\chi _i$
fields as there are $U(1)_i$ factors in the gauge group.

We will be generally interested in $\prod _{i=1}^n U(N_i)$ gauge
theories.  Chiral fermions in the fundamental $\fund _i$
contribute to \anomg\ as $[\tr \exp ({i F_i\over 2\pi})]^{(1)}$,
chiral fermions in the anti-fundamental $\overline {\fund _i}$
contribute as $[\tr \exp (-{i F_i\over 2\pi})]^{(1)}$, chiral
fermions in the adjoint do not contribute in 4d, and chiral
fermions in the $(\fund _i, \overline{\fund _j})$ contribute as $[
\tr \exp ({i F_i\over 2\pi})\tr \exp (-{iF_j\over 2\pi})]^{(1)}$,
where for all these one just keeps the 6-form term and $\tr $ is
in the fundamental of the corresponding $U(N_i)$.  The general
anomaly thus obtained is \eqn\anompg{\delta _\lambda \log Z=[\sum
_i \A _{iii}\tr F_i^3+ \sum _{ij}\A _{ijj}\tr F_i\tr
F_j^2]^{(1)}.} (No $\A _{ijk}[\tr F_i \tr F_j \tr F_k]^{(1)}$
terms occur if the only mixed representations are bi-fundamentals,
as is the case in theories with noncommutative generalizations.)

The $\A _{iii}$ term in \anompg, which is the $\Tr U(N_i)^3$
anomaly, cannot be cancelled if $N_i>1$ (or for $N_i=1$ in the
non-commutative case to be discussed).  We thus assume the matter
content is chosen such that $\A _{iii}=0$.  The remaining $\A
_{ijj}$ terms in \anompg\ are the $\Tr U(N_i)U(N_j)^2$ mixed
anomalies. These anomalies can be cancelled via $n$ fields $\chi
_i$ with couplings \eqn\anompwz{S_{GS}=\sum _{j=1}^n \int \chi _j
\tr F_j \wedge F_j,} if, under gauge transformations, the $\chi
_j$ shift as \eqn\anompsh{\delta _\lambda \chi _j=-\sum _i \A
_{ijj}\tr \lambda _i.}

For example, for the theory \stdmdl, the gauge anomaly is
\eqn\stdmdlaa{\delta _\lambda \log Z={3i\over 8\pi ^2}[\tr F_1\tr
F_2^2 -\tr F_2\tr F_1^2+\tr F_2\tr F_3^2-\tr F_3\tr F_2^2+\tr
F_3\tr F_1^2 -\tr F_1\tr F_3^2]^{(1)}.} To cancel these anomalies,
one needs three fields $\chi _i$, $i=1,2,3$, with couplings
\eqn\stdmdlci{S_{GS}=\int \sum_{i=1}^3 \chi _i \tr F_i\wedge F_i,}
which shift under gauge transformations as \eqn\stdmdlcis{\delta
\chi _1 \sim (\tr \lambda _2 - \tr \lambda _3), \quad \delta \chi
_2 \sim (\tr \lambda _3-\tr \lambda _1), \quad \delta \chi _3\sim
(\tr \lambda _1 -\tr \lambda _2).} In string theory constructions
of commutative gauge theories via branes at orbifold
singularities, the required Green-Schwarz fields $\chi _i$ arise
as Ramond-Ramond fields, coming from the closed string twisted
sector \refs{ \DouglasSW, \IbanezQP}.  They can be regarded as
e.g. the IIB $C_2$ field reduced on the collapsed $S^2$'s of the
orbifold singularity.  These are the same fields which yield the
moduli for changing the field theory theta angles of the orbifold
quiver gauge theory (see e.g. the discussion in \LawrenceJA).

\newsec{Noncommutative gauge theories and their anomalies}

Taking the $x_2$ and $x_3$ spatial directions non-commutative,
$[x_2,x_3]=i\theta _{23}$, the gauge group is generally $U(N)$,
with the $U(1)$ and $SU(N)$ parts of $U(N)$ necessarily coupled by
the noncommutative gauge symmetry. The only representations which
properly represent the noncommutative algebra are the singlet,
fundamental, anti-fundamental, and adjoint representations of
$U(N)$.  Writing the gauge transformation in terms of
$U=e^{i\lambda}$ (expanding to first order in $\lambda$ to get the
Lie algebra), the gauge field transforms as $(d+iA)\rightarrow
U\star (d+iA)\star U^{-1}$, a fundamental matter field $\psi _f$
(scalar or fermion) transforms as $\psi _f\rightarrow U\star \psi
_f$ an anti-fundamental as $\widetilde \psi _{\bar f}\rightarrow
\widetilde \psi _{\bar f}\star U^{-1}$, and an adjoint as $\psi
_a\rightarrow U\star \psi _a\star  U^{-1}$.

It is also possible to consider product gauge groups $\prod _i
U(N_i)$, in which case the only possible mixed representations are
bifundamentals $\psi _{i\bar j}$ transforming as $(\fund _i,
\overline{ \fund _j})$, i.e. $\psi _{i\bar j}\rightarrow U_i\psi
_{i\bar j}U_j ^{-1}$, with $U_i$ and $U_j$ the $U(N_i)$ and
$U(N_j)$ gauge transformations. A bi-fundamental chiral fermion
$\psi _{i\bar j, \alpha }\in (\fund _i, \overline{\fund _j})$
contributes to the $U(N_i)$ current as $J^{U(N_i)}_\mu =\sigma
^{\alpha \dot \alpha}_\mu \psi _{i\bar j, \alpha}\star \psi
^\dagger _{\bar j i, \dot \alpha}$ and to the $U(N_j)$ current as
$J^{U(N_j)}_\mu =\sigma ^{\alpha \dot \alpha}_\mu \psi ^\dagger
_{\bar j i, \dot \alpha}\star \psi _{i\bar j, \alpha}$, as these
orderings properly transform as $J^{U(N_i)}_\mu \rightarrow U_i
J^{U(N_i)}_\mu U_i^{-1}$. Other mixed representations, including
$(\fund _i , \fund _j)$, are not allowed, as they do not properly
represent the group multiplication.

\subsec{$\Tr U(N)^3$ anomalies}

These anomalies come entirely from planar diagrams.  We briefly
review the results for the non-commutative theories, as has been
analyzed e.g. in \refs{\Ardalan - \BonoraHE}. Recall first the
commutative case. Define $\Gamma _{\mu _1\mu _2\mu
_3}^{a_1a_2a_3}(p_1,p_2,p_3)= (2\pi)^4\delta (p_1+p_2+p_3)\Gamma
_{\mu _1\mu _2\mu _3}^{a_1a_2a_3}(p_1,p_2)$ to be the Fourier
transform of $\langle T J_{\mu _1}^{a_1}(x_1) J_{\mu
_2}^{a_2}(x_2) J_{\mu _3}^{a_3}(x_3)\rangle$, with $J^{a}_\mu$ the
left-handed chiral current.  The commutative anomaly \anomg\ comes
{}from \eqn\anomci{p_3^{\mu _3}\Gamma _{\mu _1\mu _2\mu
_3}^{a_1a_2a_3}(p_2,p_3)= {1\over 24\pi ^2}\epsilon _{\mu _1\mu
_2\rho \sigma} p_1^\rho p_2^\sigma \Tr _\rho
(T^{a_1}T^{a_2}T^{a_3}+T^{a_1}T^{a_3}T^{a_2}),} with $\rho$ the
representation of all left-handed chiral fermions.  The two terms
in \anomci\ come from the two contributing diagrams, one where
the cyclic ordering of external momenta is $(p_1,p_2,p_3)$ and the
other where it's $(p_2,p_1,p_3)$.  The anomaly \anomci\ vanishes
if \eqn\anomv{\Tr _\rho(\{T^{a_1},T^{a_2}\}T^{a_3}) \equiv \A
d^{abc}=0.}

The same two planar diagrams contribute in the non-commutative
case and thus the only modifications are in phase factors depending
on the external momenta: \eqn\anomnci{p_3^{\mu _3}\Gamma _{\mu
_1\mu _2\mu _3}^{a_1a_2a_3}(p_2,p_3)= {1\over 24\pi ^2}\epsilon
_{\mu _1\mu _2\rho \sigma} p_1^\rho p_2^\sigma \Tr _\rho
(e^{-{i\over 2}p_1\times p_2}T^{a_1}T^{a_2}T^{a_3}+ e^{{i\over
2}p_1\times p_2} T^{a_1}T^{a_3}T^{a_2}),} with $p_1\times
p_2\equiv p_{1\mu}\Theta ^{\mu \nu}p_{2\nu}$. As noted in \Martin,
vanishing anomaly \anomnci\ appears to require a stronger
condition than in the commutative context: \eqn\anomncv{\Tr _\rho
T^{a_1}T^{a_2}T^{a_3}=0.} However, given that the only allowed
matter representations are the fundamental, anti-fundamental, or
adjoint representation, \anomv\ and \anomncv\ actually imply
exactly the same condition, that the fermion matter content of a
single gauge group be vector-like (equal number of fundamental and
anti-fundamental chiral fermions).  Even for $U(2)$, the $U(1)^3$
part of \anomv, along with the restriction on the classically
allowed representations, would suffice to give the same
information is \anomncv, that the chiral fermions must have equal
numbers of $\fund$ and $\overline{\fund}$ chiral reps (which are
not equivalent).

In the commutative context, the anomaly \anomci\ leads to \anomg,
which satisfies the Wess-Zumino consistency conditions.  In the
non-commutative context, the claim of \BonoraHE\ is that the
anomalies can still be written via a generalization of the descent
formalism, as \eqn\anomncg{\delta _\lambda \log Z\sim \int \Tr
_\rho (\lambda \star d(A\star dA+\half A\star A\star A)).} A
subtlety here is that in the usual descent formalism,
$I_{d+2}=dI^{(0)}$ and $\delta I^{(0)}=dI^{(1)}$, the forms $I$
are local and involve traces over the gauge group, whereas in the
non-commutative context the trace over the gauge group should
include an integral over the noncommutative space.  E.g. the usual
descent formalism makes use of the cyclic property of the trace,
which wouldn't hold in the non-commutative context without
including the integral over space; it is argued in \BonoraHE\
that, nevertheless, a descent formalism can be applied and leads
to anomalies such as \anomncg.

\subsec{Mixed $\Tr U(N_i) U(N_j)^2$ anomalies}

These anomalies come entirely from non-planar diagrams, with
$U(N_i)$ gauge fields on one edge of the annulus and $U(N_j)$
gauge fields on the other.  Here the difference between the
commutative vs non-commutative cases is much more radical than for
the $\Tr U(N)_i^3$ anomalies discussed in the previous subsection.

The additional phase factors of non-planar diagrams in the
noncommutative theory depend on the loop momentum, regulating the
integrals in a way which is compatible with the naive
(non-anomalous) Ward identities.  For example, the naive Ward
identity of the commutative theory is violated because one cannot
shift the momentum integration variable of the triangle diagram,
as the integral is linearly divergent.  But the additional phase
factor of the non-planar diagrams in the non-commutative theory
leads to momentum integrals such as, quoting  \Martin,
$$\int {d^4 q\over 2\pi ^4}e^{iq\times p}{q^\mu \over q^2(q+p)^2}=
-{ip^\mu \over 8\pi ^2}\int _0^1 dx \, x K_0(\sqrt{\widetilde
p^2(-p^2) x(1-x)})$$
$$+{1\over 8\pi ^2}{\widetilde p^\mu \over \widetilde p^2}\int _0^1
dx \sqrt{\widetilde p^2(-p^2)x(1-x)}K_1(\sqrt{\widetilde
p^2(-p^2)x(1-x)}),$$ with $\widetilde p^\mu \equiv \theta ^{\mu
\nu}p_\nu$ and $\widetilde p^2 \equiv \widetilde p^\mu \widetilde
p^\nu \eta _{\mu \nu}\equiv p\circ p$.  Thus, for any non-zero
non-commutativity $\theta^{\mu \nu}$, the momentum integrals are
finite.  One can therefore freely shift the integration variable
and recover the naive Ward identities, as was done in \Martin.

The upshot is generally a radical difference from commutative
theories.  One could start with a commutative theory with non-zero
mixed anomalies (including ABJ type flavor anomalies), turn on an
arbitrarily small non-commutativity, and suddenly these anomalies
magically disappear.

\newsec{Noncommutative Green-Schwarz mechanism}

We interpret the automatic vanishing of mixed anomalies in
noncommutative gauge theories as an example of the UV/IR mixing of
non-commutative theories: the low energy theory automatically
contains additional ``closed string'' modes, even if they were
apparently not included in the original Lagrangian.  These fields
lead to automatic Green-Schwarz cancellation of all mixed
anomalies.

To motivate this claim, consider a general $\prod _{i=1}^n U(N_i)$
gauge theory whose chiral fermion matter content has non-zero
mixed $\Tr _\rho U(N_i)U(N_j)^2$ anomalies $\A _{ijj} \neq 0$ in
the commutative context.  An example is \stdmdl, with anomaly
\stdmdlaa. This means that the $U(N_j)$ instanton 't Hooft vertex
${\cal O}^{zm}_j$, associated with the $U(N_j)$ instanton's
fermion zero modes, is not a $U(N_i)$ singlet.  If $U(N_i)$ is a
gauge symmetry, the net $U(N_i)$ charge of the $U(N_j)$ instanton
must be screened, by the charged Green-Schwarz field $e^{i\chi
_j}$, in order to not spoil $U(N_j)$ gauge invariance.

In the noncommutative version of the theory, one seemingly gets no
mixed anomaly, and thus no need for a GS mechanism - but we claim
that the proper interpretation is rather that noncommutative field
theory automatically implements a noncommutative analog of the GS
mechanism. Indeed, we expect the noncommutative $U(N_j)$ instanton
to have the same fermion zero modes as in the commutative theory,
and thus the same factor ${\cal O}_j^{zm}$, with net $U(N_i)$
charge, requiring screening by some GS fields $e^{i\chi _j}$.

We claim that the zero mixed anomaly should be interpreted as a GS
cancellation of a non-zero anomaly \eqn\ncmix{\delta _\lambda \log
Z=\sum _{ij}\int d^4 k \A _{ijj}\lambda _i (k) {\cal O}_{\chi
_j}(-k),} which is the natural, non-commutative generalization of
the commutative mixed anomalies.  Here ${\cal O}_{\chi _j}(-k)$ is
the non-local, gauge invariant, analog of $\Tr F_j\wedge F_j$
which involves an open Wilson-loop: \eqn\ochidf{{\cal O}_{\chi
_j}(-k)=\Tr \int d^4 x L_*[\sqrt{\det (1-\theta \widehat
F_j)}\left(\widehat F_j {1\over 1-\theta \widehat
F_j}\right)\wedge \left(\widehat F_j{1\over 1-\theta \widehat
F_j}\right)W(x,{\cal C}_k)]\star e^{-ik\cdot x},} where we follow
the notation of \LiuMJ\ and the operator \ochidf\ appeared as the
operator $Q_4(-k)$ in \LiuPK.    The pre-cancelled anomaly \ncmix\
contains the information about the $U(N_i)$ charge of the $U(N_j)$
instanton's fermion zero modes (see e.g. the discussion of index
theorems in \LiuPK). It is natural for the non-local operator
${\cal O}_{\chi _j}(-k)$ to appear in the noncommutative mixed
anomaly \ncmix, as the mixed anomalies have non-zero non-planar
momentum and thus involve stretched string factors as in
\refs{\LiuQH, \LiuAD}.

The $\lambda _i(k)$ appearing in \ncmix\ is defined as follows.
Following the discussion in \LiuPK, we define $A_{i\mu}(k)$ by
\eqn\Ancdf{(dA_j)(k)=\Tr \int d ^4 x L_*[\sqrt{\det (1-\theta
\widehat F_j)} \left(\widehat F_j {1\over 1-\theta \widehat
F_j}\right)W(x,{\cal C}_k)]\star e^{-ik\cdot x};} this is the
Seiberg-Witten \SW\ map solution, conjectured in \LiuMJ\ and
proven in \refs{\LiuPK, \MukhiVX, \OkawaMV} by considering
Ramond-Ramond couplings (of the same kind as we're considering):
the $A_{i\mu}(k)$ defined by \Ancdf\ is simply the {\it
commutative} $U(1)_j\subset U(N_j)$.  Under a general
noncommutative $U(N_j)$ gauge transformation $\delta _{\widehat
\lambda _j}\widehat A _{j\mu}=\partial _\mu \widehat \lambda _j
+i\widehat \lambda _j \star \widehat A _{j\mu}-i \widehat A
_{j\mu}\star \lambda _j$, the $A_{j\mu}$ defined by \Ancdf\ should
then transform as $A_{j\mu} (k)\rightarrow A_{j\mu}(k) +(\partial
_\mu \lambda _j)(k)$.  This defines $\lambda _j(k)$, the gauge
variation of the commutative \Ancdf\ $U(1)_j\subset U(N_j)$, which
appears in the proposed (pre-cancelled) anomaly \ncmix.

According to the Seiberg-Witten \SW\ map, the non-commutative
theory can be mapped to a commutative version of the theory,
deformed by higher dimension (generally non-local) operators.
Because anomalies are generally independent of such deformations,
it is natural to expect the anomalies of the noncommutative theory
to be the same as that of the commutative theory.  This agrees
with the proposed pre-cancelled anomaly \ncmix: it is just the
same mixed anomaly of the commutative theory, written in the
non-commutative variables via the Seiberg-Witten map.

We propose that the anomaly \ncmix\ is automatically cancelled by
the following non-commutative generalization of the Green-Schwarz
mechanism.  For our general $\prod _{i=1}^n U(N_i)$ gauge theory,
one must introduce $n$ ``closed string'' GS fields $\chi _j$,
which are almost gauge invariant, with coupling
\eqn\ncgs{S_{GS}=\sum _{j=1}^n \int d^4 k \chi _j (k){\cal
O}_{\chi _j}(-k),} with the ${\cal O}_{\chi _j}(-k)$ the gauge
invariant operators \ochidf.  Now suppose that the gauge invariant
field congugate to $\chi_j $ is \eqn\nchj{H_{j\mu}(k)=(\partial
_\mu \chi _j) (k)+ \sum _i \A _{ijj} A_{i\mu} (k),} with $A_{i\mu}
(k)$ the commutative $U(1)_i\subset U(N_i)$ vector potential
defined via \Ancdf.  Then, the $\chi _j(k)$ cannot be precisely
gauge invariant, rather: \eqn\ncgss{\delta _\lambda \chi _j (k) =
- \sum _i \A _{ijj}\lambda _i(k).} Combining \ncgss\ and \ncgs\
cancels the anomaly \ncmix.

The coupling \ncgs\ implies that the noncommutative $U(N_j)$
instanton's 't Hooft vertex is $e^{i\chi _j(k)}{\cal
O}_j^{zm}(-k)$ and, via \ncgss, the $e^{i\chi _j(k)}$ factor
carries the appropriate charge under the commutative
$U(1)_i\subset U(N_i)$ to cancel off that of the zero mode factor
${\cal O}_j^{zm}(-k)$.

Given \nchj, the possible kinetic terms of the ``closed string''
modes $\chi _i$ are \eqn\nckin{-\half \int d^4 k \sum
_{jj'}Z_{jj'}H_{j\mu}(k)H_{j'}^\mu (-k),} for some normalization
constants $Z_{jj'}$ (which could vanish in the decoupling limit).
Despite the fact that the $U(1)$ cannot be decoupled from the
$U(N)$ in noncommutative theories, the gauge invariant terms
\nckin\ contains non-local photon mass terms like
$A_{i\mu}(k)A_{i'}^\mu (-k)$.  Via the Seiberg-Witten map, these
are just the standard, commutative photon mass terms of the GS
mechanism for the overall $U(1)_i \subset U(N_i)$.

The coupling \ncgs\ is of the same general form argued in
\refs{\MukhiVX, \OkawaMV, \LiuPK}\ to appear in the $S_{WZ}$
Wess-Zumino coupling of closed string RR fields to the worldvolume
gauge theory. Indeed, we expect that the $\chi _i$ arise in
precisely this way, from closed string twisted sector RR fields.
These can be regarded, following \LawrenceJA, as the RR field
$C_2$ reduced on the collapsed $S^2$s of the orbifold.  As
discussed in \refs{\MukhiVX, \OkawaMV, \LiuPK}, the $S_{WZ}$ terms
yield the solutions of the Seiberg-Witten map \SW\ between
noncommutative and commutative gauge theories, as $S_{WZ}$ is
actually independent of the interpolator $\Phi$ of \SW. Using the
Seiberg-Witten map (assuming it exists for our theories), \ncmix\
is simply the same anomaly of the commutative theory, written in
the noncommutative variables. Similarly, our claimed
non-commutative GS mechanism \ncgs\ and \ncgss\ is essentially the
same as the usual commutative GS mechanism, just written in the
noncommutative variables.

\subsec{Deconstructing the zero mixed anomaly}

D-brane worldvolume anomalies can be analyzed in analogy with the
discussion in \GreenMN.  Consider, in particular, D3 brane
worldvolume anomalies when at orbifold singularities, such as the
$C^3/Z_3$ example.  The mixed $\Tr U(N_i)U(N_j)^2$ anomaly comes
from the amplitude with a $U(N_i)$ gauge field on one boundary of
the annulus and two $U(N_j)$ gauge fields on the other boundary.
There is an anomaly if this amplitude does not vanish when the
$U(N_i)$ polarization is set equal to its momentum $k_{3\mu}$. We
regulate the amplitude as in \GreenMN\ by introducing a
Pauli-Villars field of mass $M$ and then take $M\rightarrow
\infty$.  The anomalous amplitude is then proportional to
\eqn\anomamp{  A_{ST}=\lim _{M\rightarrow \infty}A_{FT} M^2 \int
_0^\infty dt e^{-M^2t} e^{-{\pi \over 2t}\alpha ' k_{3\mu} g^{\mu
\nu}k_{3\nu}}\int _0^1 d \nu  _{1,2}e ^{-{i\over 2}(k_1\times
k_2)[2\nu  _{12}+\epsilon (\nu _{12})]},} where $A_{FT}\equiv \Tr
_\rho (U(N_i)U(N_j)^2)k_{1\mu}k_{2\nu}\epsilon ^a_{1\rho}\epsilon
^a _{2\sigma}\epsilon ^{\mu \nu \rho \sigma}$ is the standard
commutative field theory amplitude for the anomaly.  In the
notation of \GreenMN, $q=\exp (-2\pi /t)$ and \anomamp\ contains
the characteristic non-planar factor $q^{-s/4}$, where $s\equiv
-\alpha ' k_{3\mu}g^{\mu \nu}k_{3\nu}$ is the non-planar
$s$-channel invariant momentum (with $s<0$ as defined here).  Note
that $s$ involves only the closed string metric $g^{\mu \nu}$
(rather than the open string metric $G^{\mu \nu}$) for any
non-commutivity.

Setting $\theta ^{\mu \nu}=0$ and $\alpha '=0$, the amplitude
\anomamp\ gives $A_{ST}=A_{FT}$.  For string theory, with $\alpha
'\neq 0$, \anomamp\ gets a factor $\sim \lim _{M\rightarrow
\infty} e^{-\alpha ' M^2}\rightarrow 0$, thus there is no anomaly.
This is the usual statement that string theory cancels this mixed
anomaly via closed string modes (twisted sector RR fields) and the
Green-Schwarz mechanism.  The factor $q^{-s/4}$ corresponds to the
closed string poles in the non-planar $s$-channel.

Now in the noncommutative field theory limit we take $\alpha
'\rightarrow 0$ and adjust the closed string metric $g^{\mu \nu}$
so that the open string $G^{\mu \nu}=\eta ^{\mu \nu}$ is the
identity; this gives $\alpha ' p_\mu g^{\mu \nu}q_\nu \rightarrow
p_\mu (\theta ^2)^{\mu \nu}q_\nu \equiv p\circ q$ \MinwallaPX.
Thus the anomaly amplitude \anomamp\ becomes \eqn\anomampnc{
A_{NCFT}=\lim _{M\rightarrow \infty}A_{FT} M^2 \int _0^\infty dt
e^{-M^2t} e^{-{\pi \over 2t} k_3 \circ k_3} \int _0^1 d \nu
_{1,2}e ^{-{i\over 2}(k_1\times k_2)[2\nu  _{12}+\epsilon (\nu
_{12})]},} which still gives zero for the anomaly,
$A_{NCFT}\rightarrow 0$ in the $M\rightarrow \infty$ limit, a
result which we have already discussed. Since the vanishing
anomaly comes from the same $q^{-s/4}$ closed string factor as in
the commutative string theory version of the theory, it naturally
has the same interpretation: a non-zero anomaly is cancelled by
closed string twisted sector RR states $\chi _j$.

We would now like to deconstruct the zero anomaly of \anomampnc\
to see that it corresponds to an automatic GS cancellation of the
proposed anomaly \ncmix.  Suppose we drop the ``closed string''
factor $q^{-s/4}\equiv e^{-{\pi \over 2t} k_3 \circ k_3}$ in
\anomampnc, which is responsible for the anomaly cancellation.
Then the remaining parts of the anomaly in \anomampnc\ would give
a non-zero anomaly, which differs from the commutative result by
the extra phase factors coming {}from the $\nu _{1,2}$ integrals.
These phase factors lead to $\star '$ products as in
\refs{\GarousiCH, \LiuAD}, and the resulting anomaly is of the
form
$$
\partial^{\mu} \tr J^{U(N_i)}_{\mu} \propto \tr
\hat F ^{U(N_j)}\wedge_{\star '} \hat F^{U(N_j);}
$$
which is a leading term corresponding to the claimed mixed anomaly
\ncmix.  It should be possible to obtain the full gauge invariant
operator corresponding to \ncmix\ from a one-loop computation in
some properly interpreted sense (presumably, properly interpreted,
the Adler-Bardeen theorem applies).

\subsec{$U(1)^3$ anomalies cannot be cancelled}

In the commutative case, $U(1)$ is qualitatively different from
$U(N>1)$, in that non-zero $\Tr U(1)^3$ anomalies of a chiral
$U(1)$ theory can be cancelled via the GS mechanism as in \pff,
whereas $\Tr U(N>1)^3$ anomalies cannot (because of the $\Tr
SU(N)^3$ part).  In noncommutative theories, $U(1)$ is Morita
equivalent to $U(N>1)$, so one should expect that GS anomaly
cancelation either works or fails for cancelling $\Tr U(N)^3$
anomalies for all $N$.  We claim that, unlike the commutative
case, there is no noncommutative GS mechanism to cancel nonzero
$\Tr U(1)^3$ anomalies.

Consider a noncommutative chiral $U(N)$ theory, with non-vanishing
$\Tr U(N)^3$ anomaly coming from the planar diagrams with three
$U(N)$ gauge fields on a single boundary of the annulus. Because
these diagrams are planar, they differ from the commutative theory
only by the phase factors depending on external momenta,
corresponding to the replacement of ordinary products with $\star$
products, leading to \anomncg. The question is whether or not
these loop diagrams can be cancelled by a tree-level diagram
involving a field $\chi$ with a noncommutative analog of the
interaction \pff.

But the only plausible noncommutative analog of the GS mechanism
is to have the GS field $\chi$ be a gauge singlet (up to its
anomalous shift).  In particular, $\chi$ should not be an adjoint
field under the gauge group, so the only plausible noncommutative
analog of \pff\ is as in \ncgs: \eqn\pffnc{S_{GS}=\int d^4 k \chi
(k){\cal O}_\chi (-k),} with ${\cal O}_\chi (-k)$ the gauge
invariant version of $\Tr F\wedge F$, defined as in \ochidf.  This
could only cancel an anomaly like \eqn\anomcc{\delta _\lambda \log
Z \sim \int \lambda (k){\cal O}_\chi (-k),} which is a possible,
but incorrect, noncommutative analog of \anomvi. It must be
incorrect because the $\Tr U(N)^3$ anomaly comes from a planar
diagram and yields the anomaly discussed above.  Anomalies like
\anomcc, involving the gauge invariant operators ${\cal O}_\chi
(-k)$, can only come from non-planar diagrams.  We can explicitly
see that \anomcc\ does not agree with \anomncg\ by noting that the
leading order term in ${\cal O}_\chi$ is $F\wedge _{\star '} F$,
and $d[A\star dA -{1\over 2}A\star A\star A] \neq F\star ' F$.

Our conclusion is that non-zero non-commutative $\Tr U(1)^3$
anomalies cannot be cancelled by any obvious generalization of the
Green-Schwarz type mechanism.  The only way then to avoid $\Tr
U(1)^3$ anomalies is to have non-chiral fermion representations,
as is also the case for $U(N>1)$ already in the commutative
theory.

This conclusion would be problematic if one could find a string
theory construction which yields a chiral $U(1)$ gauge theory,
with non-zero $\Tr U(1)^3$ anomaly cancelled by the commutative GS
mechanism, which admits turning on $B_{NS}\neq 0$.  Fortunately,
consistent string theory constructions, which admit turning on
$B_{NS}\neq 0$, do not need to make use of the commutative GS
mechanism for cancelling $\Tr U(1)^3$ anomalies, as the individual
$U(1)$ groups are generally nonchiral, with $\Tr U(1)^3=0$.

\newsec{Comments on noncommutative Higgsed theories,
like the noncommutative standard model}

Consider a noncommutative analog of the standard model (ignoring
hypercharge), with gauge group $U(3)\times U(2)$ and with each
generation's chiral fermion matter content as in
\eqn\stdmdl{\matrix{ \quad &U(3)\times U(2)\cr q&(\fund ,
\overline{\fund})\cr \overline u&(\overline{\fund}, 1)\cr
\overline d&(\overline{\fund} , 1)\cr l&(1,\fund )}} The anomaly
is given by descent from the 6-form term in \eqn\anomix{\tr \exp
(F_3/2\pi)\tr \exp(-F_2/2\pi)+ 2\tr \exp(-F_3/2\pi)+\tr
\exp(F_2/2\pi),} i.e. \eqn\anomixi{\delta_\lambda \log Z={i\over
8\pi ^2}[\tr F_3\tr F_2^2- \tr F_2\tr F_3^2 -{2\over 3}\tr F_2
^3]^{(1)}.}  In the commutative case these anomalies all involve
only $U(1)$ factors (as $SU(2)$ has no cubic Casimir) and can be
cancelled by introducing two GS fields $\chi _3$ and $\chi _2$.

In the non-commutative case, the mixed anomalies in \anomix\ are
cancelled automatically, but the $\Tr U(2)^3$ anomaly in \anomix\
cannot be GS cancelled, as discussed above. The theory with the
matter content \stdmdl\ would be incurably sick, because its
$U(2)$ matter content is not vector-like.  However, because the
$U(2)$ is broken anyway by the Higgs mechanism, it is possible to
include additional $U(2)$ chiral fermions such that the total
matter content is vector-like. E.g. in the supersymmetric analog
of the above theory, one has Higgs superfields $H_u$ and $H_d$ in
the $\fund $ of $U(2)$, with Yukawa couplings like $u\bar q H_u$.
Adding these two fields to those of a single generation of
\stdmdl\ makes the fermion content vector-like.  Even with
additional generations, one could always add additional $U(2)$
fundamental matter to make the net matter content vector-like.
These additional chiral fields can be get large masses via making
their expectation values or Yukawa couplings large.

\lref\WittenTW{ E.~Witten,
Nucl.\ Phys.\ B {\bf 223}, 422 (1983).
} Chiral fermions which can contribute to the anomalies can only
be massive in Higgsed theories, with the masses arising via Yukawa
couplings.  As usual, integrating out such massive chiral fermions
of Higgsed theories leads to Wess-Zumino-Witten terms, which is
how their contribution to anomalies shows up in the low energy
effective field theory \WittenTW.  In the noncommutative theory,
the relevant diagrams are planar, so the same WZW terms are
generated (e.g. the lengthy one in eqn. (24) of \WittenTW), with
the only difference being that all products are replaced with
$\star$ products.

\lref\NishimuraDQ{ J.~Nishimura and M.~A.~Vazquez-Mozo,
hep-th/0107110.
}
\medskip

{\bf Note added:} A lattice construction of noncommutative
theories with adjoint chiral fermions, for arbitrary even
dimensions, was recently reported in \NishimuraDQ.  In $d=2$ mod 4
dimensions, the adjoint chiral fermion is not vector-like and
would contribute to anomalies in the commutative context.
Surprisingly, no anomaly contribution was found in the lattice
construction of \NishimuraDQ\ for any $d$ (though such anomalies
generally do not vanish in the continuum version of these
noncommutative theories, as they come from planar diagrams).  This
is perhaps suggestive of a lattice noncommutative cancellation
mechanism similar to that discussed here, and we thank one of the
authors of \NishimuraDQ\ for bringing this to our attention.
However, because the missing anomalies are generally irreducible
planar anomalies, they cannot be cancelled by the GS mechanism or
any known counter-term. So it's unclear how the lattice
noncommutative theory is managing to cancel the anomaly.

\bigskip
{\bf Acknowledgements}

We are grateful to S. Hellerman, A. Manohar, J. Michelson, S.-J.
Rey, J. Schwarz and W. Taylor for useful discussions.  This work
is supported by DOE-FG03-97ER40546.

\listrefs

\end